\documentclass[notitlepage,10 pt]{iopart}
\usepackage{url,graphicx, caption,braket,mathdots, iopams}

\begin{document}

\def\Gv{\mathbf{G}}
\def\Iv{\mathbf{I}}
\def\Sv{\mathbf{S}}
\def\Fv{\mathbf{F}}
\def\Nv{\mathbf{N}}
\def\Rv{\mathbf{R}}
\def\Jv{\mathbf{J}}
\def\Lv{\mathbf{L}}
\newcommand{\Dstate}[0]{\ensuremath{A'^{2}\Delta_{3/2}}}
\newcommand{\Pstate}[1]{$A^{2}\Pi_{#1/2}$} %give it a 1 or 3 for Omega = 1/2 or 3/2
\newcommand{\Sstate}[0]{\ensuremath{X^{2}\Sigma}}

\title{Prospects for a Narrow Line MOT in YO}
\author{Alejandra L. Collopy, Matthew T. Hummon, Mark Yeo, Bo Yan and Jun Ye}
\address{JILA, National Institute of Standards and Technology and University of Colorado,
and Department of Physics, University of Colorado,
Boulder, CO 80309, USA.}
\ead{collopy@jila.colorado.edu}
\begin{abstract}
In addition to being suitable for laser cooling and trapping in a magneto-optical trap (MOT) using a relatively broad ($\sim$5~MHz) transition, the molecule YO possesses a narrow-line transition. This forbidden transition between the \Sstate{} and \Dstate{} states has linewidth $\sim$2$\pi\times$160 kHz. After cooling in a MOT on the \Sstate{} to \Pstate{1} (orange) transition, the narrow (red) transition can be used to further cool the sample, requiring only minimal additions to the first stage system. The narrow line cooling stage will bring the temperature from $\sim$1~mK to $\sim$10~$\mu$K,  significantly advancing the frontier on direct cooling achievable for molecules.
\end{abstract}


\section{Introduction}
Cooling of molecules has wide-ranging applications, from tests of fundamental symmetries \cite{ACME2014,Loh2013,Hinds2011}, study of long range anisotropic interactions \cite{Yan2013}, to cold chemistry \cite{Ospelkaus2010, Bell2009}. The more complex structure of molecules as compared to atoms, though it provides a rich field of study, also makes difficult the cooling methods readily available to atoms. Indirect methods have so far produced the coldest molecules, via magneto-association and adiabatic transfer \cite{Ni2008}. Unfortunately, the number of species that can be produced in this way is limited, and so we must look to direct cooling methods to support a wider range of molecular species. Cold molecular beams can be produced via supersonic expansion \cite{Smalley1977} or from a cryogenic buffer gas beam source \cite{Hutzler2012}. Such beams can be decelerated with techniques such as Stark deceleration \cite{Fabrikant2014, Meerakker2006}, Zeeman deceleration \cite{Narevicius2008}, or via centrifugal deceleration \cite{Chervenkov2014}.  From a low velocity beam, a two photon process has been used to directly load a magnetic trap \cite{Lu2014}. Once trapped, molecular collisions and evaporative cooling have also been demonstrated for the hydroxyl radical \cite{Stuhl2012}. 

For a long time, it was thought that laser cooling of molecules was impractical due to the large number of states and therefore lasers required to implement an optical cycling transition capable of many photon scatters. However, in 2004 it was proposed \cite{DiRosa2004} that a large class of molecules has sufficently diagonal Franck-Condon factors to be amenable to laser cooling. In 2008 the first concrete proposal was made for a class of cycling transitions that through angular selection rules avoids all dark states for molecules, enabling quasi-closed cycling and so the scattering of many photons \cite{Stuhl2008}. 
In the intervening time, optical cycling has been demonstrated in a number of diatomic species, leading to demonstration of laser cooling \cite{Shuman2010}, radiation pressure slowing \cite{Barry2012,Zhelyazkova2014,Yeo2015}, and magneto-optical trapping in both two \cite{Hummon2013} and three dimensions \cite{Barry2014,McCarron2014}.  Additionally, it has been proposed that the optical bichromatic force \cite{Eyler2011} or ultrafast stimulated slowing \cite{Jayich2014} can be used for molecular beam deceleration and cooling \cite{Metcalf2008}. 

The lowest temperature realized with these direct cooling techniques is so far that of the magneto-optical trap (MOT) \cite{Barry2014}, with temperature $\sim$2~mK. While a short-lived excited state is desirable for quickly slowing a molecular beam, in order to decrease the final temperature of the MOT a narrower linewidth transition is needed \cite{Katori1999,Loftus2004}. 
 After careful study of the level structure of YO for the primary cooling transition, we identify a transition in YO that will enable narrow line cooling to the Doppler temperature $T_{D} =\hbar\gamma/2k_{B}$ $\approx$ 5~$\mu$K where $\hbar$ is the reduced Planck constant, $\gamma$ is the linewidth of the transition,  and $k_{B}$ Boltzmann's constant. Our proposed scheme will only require the addition of one laser to the primary cooling setup of the YO molecule. Narrow line cooling will greatly decrease the temperature limit of direct molecular cooling processes, and allow the loading of molecules into an optical dipole trap for possible evaporative cooling \cite{Grimm2000}. In this paper, we estimate the lifetime of the \Dstate{} state and calculate its vibrational branching rates in order to characterize the transition and assess its feasibility for use in a MOT. Additionally, we detail the cycling transition that will be used to create a narrow-line MOT and simulate its effects and cooling trajectory.

\section{Cooling process}
\begin{figure}[h]
\includegraphics[width=\textwidth]{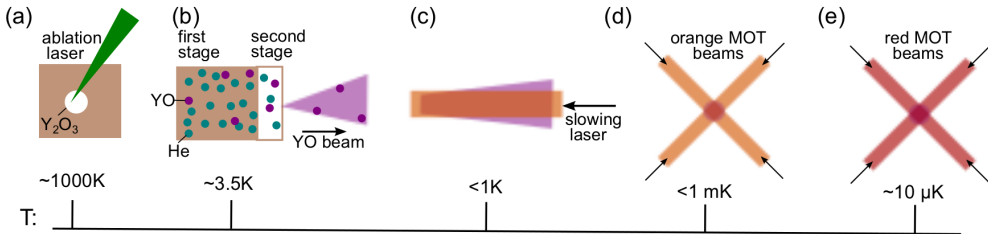}
\caption{Cooling process schematic. After production via ablation (a), YO molecules thermalize with helium at 3.5~K in a two stage buffer gas cell (b). Upon exiting, the beam has velocity $\sim$70~m/s. Longitudinal white light slowing produces molecules \textless10~m/s (c), loadable into a MOT. The orange MOT will produce cold molecules less than 1~mK (d). Narrow line cooling can then be performed on the red transition (e).}
\label{fig:schematic}
\end{figure}

The entire process to cool YO from initial production down to the $\mu$K scale involves a number of steps, as depicted in Figure \ref{fig:schematic}. We form YO via laser ablation of a Y$_{2}$O$_{3}$ target.
The initial temperature after ablation is high both rotationally and translationally, whereas any optical cycling scheme requires a large number of molecules in a single state. To meet this requirement, we utilize a two-stage cryogenic buffer gas cell \cite{Lu2011} to produce cold YO molecules. The buffer gas cell is anchored to a 3.5~K cold plate and helium gas is constantly flowed in to maintain a high density ($\sim10^{16}$/cm$^{-3}$) of helium in the cell. The helium collisions with the cell walls cool the helium, and YO collisions with the helium buffer gas then rotationally and translationally cool the YO molecules. The second stage of the cell has sufficiently low helium density to ameliorate boosting of the YO beam due to collisions with the helium gas at the aperture. The net result is a beam of YO molecules with forward velocity 70~m/s with a 40~m/s full width at half maximum, corresponding to a $\sim$7~K longitudinal temperature. Longitudinal slowing of the YO beam is accomplished \cite{Yeo2015} with a combination of frequency chirping and white light modulation of the counter-propagating slowing laser, resulting in a population of YO molecules with speeds under 10~m/s, which are sufficiently slow to load into a first stage (orange) MOT. Further cooling can be accomplished with a narrow-line MOT, analagous to methods demonstrated for several atomic species \cite{Loftus2004,Loftus2004A,Berglund2008,Duarte2011,Villwock2011,Lu2011Dy,Frisch2012}.

\section{Level structure of YO}

\begin{figure}[h]
    \includegraphics[width=\textwidth]{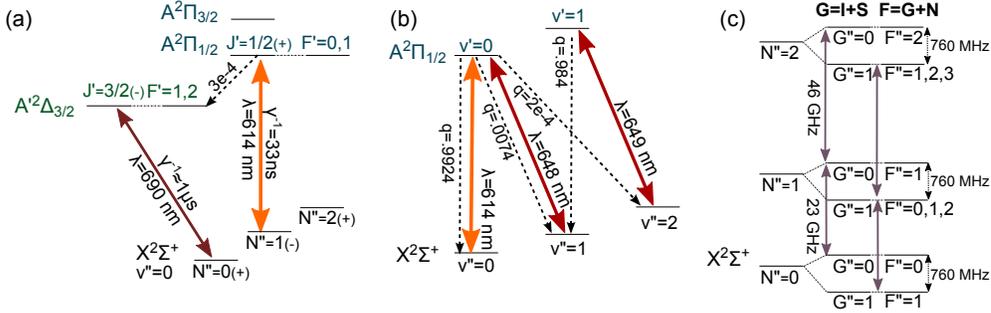}
  \caption{Relevant level structure of the YO molecule. (a) Electronic structure and ground state rotational structure. Solid lines indicate the broad (orange) and narrow (maroon) transitions that will be used to make MOTs in YO.  (b) Vibrational structure and repumping scheme. Solid lines indicate lasers for first stage cooling (orange) and vibrational repumps (red). Dashed lines show spontaneous decays and are labeled with Franck-Condon factors  (c) Ground state rotational and hyperfine structure.  For each $G$ level, $F$ ranges from $|G-N|$ to $|G+N|$. }
\label{fig:levels}
\end{figure}

The molecule YO has three electronic states that are of primary concern, the \Sstate{} ground state, the \Pstate{1} excited state, and the \Dstate{} intermediate state, as shown in Figure \ref{fig:levels}(a). For describing the angular momentum coupling in these states we follow the notation used by Brown and Carrington \cite{Brown2003}. When discussing a specific value of angular momentum, the factor of $\hbar$ is omitted.
For the states we consider in YO, we have electronic spin $S$ = 1/2 and nuclear spin $I$= 1/2. Electron orbital angular momentum is labeled $\Lv$, while rotational and total angular momenta are labeled $\Rv$ and $\Fv$ respectively. Additionally, we use intermediate angular momenta to describe the states:  $\Jv = \Fv - \Iv$, and $\Nv = \Jv-\Sv=\bf{R+L}$.  Parity of states is indicated by (+) or (-). As well as possessing an array of rotational levels, each state also has a ladder of vibrational levels labeled with quantum number $v$.

We begin by considering the electronic ground state \Sstate{}, which can be described by Hund's case (b$_{\beta S}$).  The effective Hamiltonian that describes the coupling is:
\begin{equation*}
H_\mathrm{eff} = \gamma \Sv \cdot \Nv + b \Iv \cdot \Sv + c I_z S_z
\end{equation*}
where $\gamma = -9.2$~MHZ, b = $-763${~MHz}, and $c = -28$~MHz,  representing the spin-rotation, Fermi contact, and electron-nuclear spin dipolar interaction respectively \cite{Childs1988}. The strong Fermi contact interaction couples the electron and nuclear spin to form intermediate angular momentum $\Gv = \Iv+\Sv$.  The total angular momentum is then formed by the additional coupling of $\Nv$ through the spin-rotation interaction yielding $\Fv = \Gv + \Nv$.  We also note for $\Sigma$ states, orbital angular momentum $\Lv$ is zero and so $\bf{N}={R}$. For $N$\textgreater0, levels can be expressed as a superposition of $J$ levels $J$=$N\pm1/2$ ($J$=1/2 for $N$=0).

 The excited \Pstate{1} and \Dstate{} states can both be described by Hund's case (a).  The rotational states are labeled by angular momenta $\Jv$ and $\Fv$, though the hyperfine interaction in these states is small and is not resolvable via optical transitions with the $\Sstate$. The quantum number $\Omega$ is given by the projection of $\bf{L+S}$ onto the internuclear axis. For $\Pi$ states, $L$=1, while for $\Delta$ states $L$=2. So, for the $A^{2}\Pi$ states, $\Omega$ is 1/2 or 3/2, and for $A'^{2}\Delta$ $\Omega$ is 3/2 or 5/2. The value of $\Omega$ is indicated by the subscript on the molecular term symbol, e.g. $\Dstate$ has $\Omega$ =3/2. Additionally, the two senses of projections of $\Lv$ onto the internuclear axis give rise to a pair of states with opposite parity for each $J$ level. For the \Pstate{1} state, the lambda doubling term p is 4.5072 GHz, indicating a splitting of 4.5 GHz for the opposite parity levels in the $J=1/2$ level \cite{Bernard1983}. In the \Dstate{} state, the lambda doubling is expected to be small, and has so far not been measured.

The \Sstate{} to \Pstate{1} transition ($v"$=0 to $v'$=0) is at 614~nm, as depicted in Figure \ref{fig:levels}(a). While \Pstate{1} primarily decays back to \Sstate{}, it has a $\sim$3$\times10^{-4}$ chance of decaying to \Dstate{} \cite{Yeo2015}. As will be discussed in more detail later, while the \Dstate{} to \Sstate{} transition is electric dipole forbidden, it does in fact occur due to mixing of the the nearby \Pstate{3} state with the \Dstate{} state \cite{Chalek1976}. This narrow transition has a wavelength of 690~nm.

\section{First stage laser cooling}
Our first stage of cooling has a main cycling transition at 614~nm from \Sstate{}, $N"$=1 to \Pstate{1}, $J$=1/2, as shown in Figure \ref{fig:levels}(a). Optical cycling has been demonstrated on this transition as a one and two dimensional MOT \cite{Hummon2013}, and for molecular beam slowing \cite{Yeo2015}. Decays to rotational levels other than $N"$=1 are forbidden by parity and angular momentum selection rules. We address vibrational branching by repumping the $v"$=1 level directly to the \Pstate{1}, $v'$=0 level using 648~nm light (Figure \ref{fig:levels}(b)). The $v"$=2 level is repumped using a two step process, first exciting it to \Pstate{1}, $v'$=1 at 649~nm, which predominantly decays to the $v"$=1 level, and then allowing the $v"$=1 repump light to take the molecule back to the $v"$=0 level.
 Molecules that decay to \Sstate{} via \Dstate{} end in the $N"$=0 or 2 levels because of parity and angular momentum rules. To maintain cycling, we apply microwave mixing resonant with the $N"$=0$\leftrightarrow$$N"$=1 and the $N"$=1$\leftrightarrow$$N"$=2 rotational transitions, depicted as gray arrows in Figure \ref{fig:levels}(c). In total, we estimate that the setup yields a closed cycling transition up to the $\sim10^{-6}$ level of enclosure, sufficient to slow longitudinally and trap a subset of the molecules. 

In addition to being able to perform Doppler cooling on the \Sstate{} to \Pstate{1} transition, the $G"$=1: $F"$=1,2 levels are subject to Zeeman splitting in an applied magnetic field, allowing the use of a MOT. For this transition, the greater multiplicity of states in the ground state means that coherent dark states can form \cite{Berkeland2002}, limiting the total scattering rate. In our setup, we avoid this issue by switching our laser polarization and the magnetic field synchronously at a rate approximately equal to the transition linewidth, eliminating long-lasting dark states.
Type-II MOTs like this one (operating from $F"$ to $F'=F"-1$) tend to be weaker and larger than their type-I counterparts (operating from $F"$ to $F'=F"+1$), with sizes on the order of a mm \cite{Prentiss1988}.  We anticipate the final temperature of the orange MOT to be on the order of 1~mK.  This is larger than the Doppler temperature $T_{D} = 110 \mu K$, but typically type-II MOTs do not reach the Doppler limit \cite{Tiwari2008, Barry2014}. However, the rapid polarization and magnetic field switching should negate the weakness of the type-II MOT in our case \cite{Tarbutt2014} and may allow a lower final temperature of the orange MOT. 

\section{Spectroscopy of the \Dstate{} state}
In order to characterize the properties of a second stage red MOT on the \Sstate{} to \Dstate{} transition, we first need an estimate of the linewidth of the transition. We acquire this estimate by comparing the absorption strength of the red transition to that of the known orange transition. To that end, we have identified the  $N"$=0,1,2:$G"$=1 to \Dstate{} transitions using absorption spectroscopy in our cryogenic buffer gas cell, as shown in Table \ref{table:deltastates}, in agreement with previous work \cite{Simard1992}.
\begin{indented}
\item{}
\Table{\label{table:deltastates}Identification of \Sstate{} to \Dstate{} transitions}
\hline
Transition & Frequency (THz) \\
\hline
N"=0,G"=1 $\rightarrow$  \Dstate{}(-) & 434.70310(3) \\
N"=1.G"=1 $\rightarrow$ \Dstate{}(+)& 434.67989(3)  \\
N"=2,G"=1 $\rightarrow$  \Dstate{}(-) & 434.63333(3)  \\
\hline
\endTable
\end{indented}

The energy splitting between $N"$=0$\rightarrow$ \Dstate{} and $N"$=2 $\rightarrow$  \Dstate{} is $N"(N"+1)B_{0}$, with $N"$ = 2, where $B_{0}$ is the rotational constant for \Sstate{}, $v"$=0. Evaluating this yields $B_{0}$ = 11.61 GHz, which agrees with the known rotational splitting \cite{Chalek1976}. By comparison of the absorption on the $N"$=1 $\rightarrow$ \Dstate{} transition with the $N"$=1 $\rightarrow$\Pstate{1} transition and the known 33~ns lifetime of the \Pstate{1} state\cite{Liu1977}, we estimate that the lifetime of the \Dstate{} state is $\tau= $425~ns. We note that the $N"$ = 0,2 to \Dstate{} transitions address a different parity level than the $N"$=1 to \Dstate{} transition, but we do not find a different value of $B$, indicating that the parity splitting of the \Dstate{} state is smaller than the Doppler broadened linewidth of $\sim$ 60 MHz inside the cell.

\section{Mixing of the \Dstate{} and $A^{2}\Pi$ states}

In addition to performing an absorption measurement, we also calculate a lifetime estimate. While electric dipole forbidden, the decay from the \Dstate{} state is almost entirely due to mixing with the \Pstate{3} state, on the order of 3.5\% \cite{Chalek1976}. The molecular potential can be approximated using the Rydberg-Klein-Rees (RKR) \cite{Rees1947} method for the states of interest. The curves for the \Sstate{} and $A^{2}\Pi$ states can be corroborated with previous work \cite{Sriramachandran2011}. After using a finite difference method to determine the wavefunctions to evaluate the vibrational overlap integrals and the $\Bra{v}1/r^{2}\Ket{v'}$ integrals we use the procedure outlined by Chalek and Gole \cite{Chalek1976} to compose the Hamiltonian including the electronic, rotational, vibrational, spin orbit, and coriolis operators. By diagonalizing this Hamiltonian for $J$=3/2, we express the  \Dstate{} eigenstate in terms of the Born-Oppenheimer states, the first few terms of which are:

\begin{eqnarray*}
\fl\Ket{\Psi_{\Delta_{3/2},v'=0}} \approx  0.9865\Ket{\Delta_{3/2},v'=0}-0.1582\Ket{\Pi_{3/2},v'=0}\\-0.0369\Ket{\Pi_{3/2},v'=1}+0.0176\Ket{\Pi_{1/2},v'=0}+...\\
\end{eqnarray*}

These coefficients squared indicate the mixing of the \Dstate{},$v'$=0 state with other states, with the largest being a 2.5\% mixing with the \Pstate{3}, $v'$=0 state. Summing over the various Born-Oppenheimer states and their associated calculated Franck-Condon factors yields estimates for the vibrational branching ratios for the \Dstate{} to \Sstate{} transition. These and other state branching ratios are compiled in Table \ref{table:fc}.

\begin{table}
\caption{\label{table:fc}Vibrational branching rates for various transitions of interest.}
\scriptsize
\begin{tabular}{llllllllllll}

\br
State 1&State 2&$v_{1}$ &$v_{2}$ & $q_{v_{1}v_{2}}$& &State 1&State 2&$v_{1}$ &$v_{2}$ & $q_{v_{1}v_{2}}$     \\
\mr
\Pstate{1}&\Sstate{}&0&0&0.9924 &&\Pstate{1}&\Sstate{}&1&0&0.0076\\
\Pstate{1}&\Sstate{}&0&1&0.0074 &&\Pstate{1}&\Sstate{}&1&1&0.9752\\
\Pstate{1}&\Sstate{}&0&2&2.4$\times10^{-4}$&&\Pstate{1}&\Sstate{}&1&2&0.0164\\
\Pstate{1}&\Sstate{}&0&3&2$\times10^{-6}$&&\Pstate{1}&\Sstate{}&1&3&7.8$\times10^{-4}$\\
\Pstate{3}&\Dstate{}&0&0&0.9078 &&\Dstate{}&\Sstate{}&0&0&0.9422\\
\Pstate{3}&\Dstate{}&0&1&0.0860 &&\Dstate{}&\Sstate{}&0&1&0.0568\\
\Pstate{3}&\Dstate{}&0&2&0.0059 &&\Dstate{}&\Sstate{}&0&2&0.0011\\
\Pstate{3}&\Dstate{}&0&3&3.3$\times10^{-4}$&&\Dstate{}&\Sstate{}&0&3&\textless10$^{-4}$\\
\br

\end{tabular}
\end{table}

Using 2.5\% mixing and the known $\sim$30~ns lifetime of the \Pstate{3} state \cite{Liu1977}, we estimate the lifetime of the \Dstate{} state as 1200~ns. These two estimates of the lifetime of the \Dstate{} state, 425~ns from the absorption measurement and 1200 ns from calculation are of similar order and give us a starting point for further calculations. From here on, we use an intermediate estimate of $\tau$= 1~$\mu$s, yielding a linewidth of $\gamma=2\pi\times$160~kHz.

\section{ Narrow-line cooling on the \Sstate{} to \Dstate{} transition}

\begin{figure}[h]
    \includegraphics[width=\textwidth]{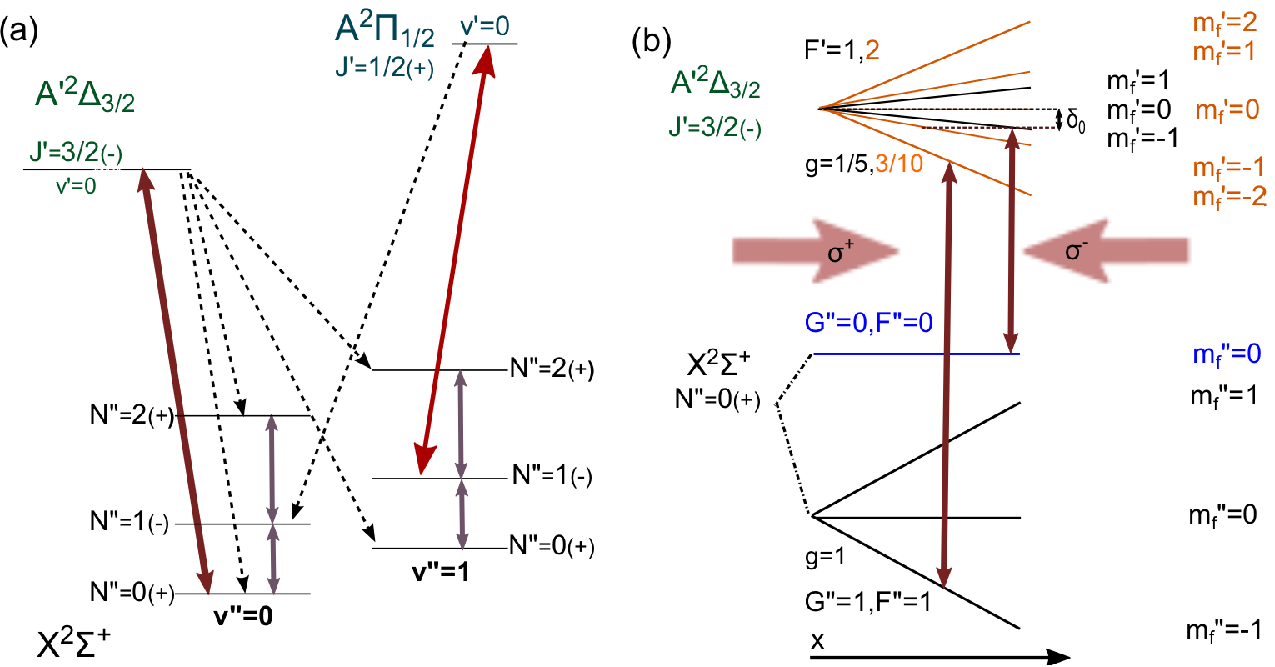}
  \caption{Level structure for the narrow-line MOT. (a) Maroon arrow indicates the transition used in the red MOT. Gray lines indicate microwave mixing which closes rotational branching. Red line indicates $v"$=1 repump, the same as used in the orange MOT. (b) Zeeman structure and MOT scheme. g represents the corresponding magnetic Land\'{e} factor. Large arrows indicate counter propagating circularly polarized laser beams. Both $F"$=0 $\rightarrow$ $F'$=1 and $F"$=1$\rightarrow$$F'$=2 transitions can effect a MOT action (maroon arrows).}
\label{fig:mot}
\end{figure}

Once molecules have been cooled with the orange MOT, we wish to further cool them utilizing the narrow \Sstate{} to \Dstate{} transition. This is acheivable by exciting from the $N"=0$ level to the $\Dstate$, $J'$=3/2 level, as in Figure \ref{fig:mot}(a). Decay from \Dstate{} ends in either $N"$=0 or 2. To maintain cycling, we use microwave mixing to address the $v"$=0, $N"$=0$\leftrightarrow$1, and $N"$=1$\leftrightarrow$2 rotational transitions, with frequencies 2$B_{0}$ and 4$B_{0}$.
Similarly, higher vibrational levels can be mixed by applying microwaves resonant with the $N"$=0$\leftrightarrow$1 and $N"$=1$\leftrightarrow$2 rotational transitions of that vibrational level, e.g. at 2$B_{1}$ and 4$B_{1}$ for $v"$=1.  Additionally, this microwave mixing scheme allows $v"$\textgreater0 vibrational levels to be repumped with the same lasers used in the orange MOT. The rotational constants for the first few vibrational levels have been determined and so mixing will be straightforward to implement \cite{Suenram1990,Bernard1979}.
The saturation intensity, $I_{sat} = (\pi hc)/(3\lambda^{3}\tau)$, is 60~$\mu$W/cm$^{2}$, which is obtainable with commercially available diodes.

In addition to the possibility of an optical molasses based on this transition, it is apparent that a narrow-line MOT is also achievable. A MOT can be made using any or all $F"$ states due to a non-zero g-factor for both the ground and excited states, but the number of frequency components and polarization considerations go up considerably with the number of states \cite{Tarbutt2014}. The simplest possible MOT action in this system is for the transition from the $N"$=0 level to the \Dstate{} state.  The $N"$=0 level has $G$ values 0 and 1, split by $\sim$760~MHz.  A type-I MOT can be performed on both sublevels, from $G"$=$F"$=1 to $F'$=2 and from $G"$=$F"$=0 to $F'$=1, as depicted in Figure \ref{fig:mot}(b). Additionally, since in type-I MOTs no Zeeman dark states can form,  experimental implementation is simplified because it will not be necessary to perform the magnetic field switching and polarization chopping used in the orange MOT.

\section{Narrow-line MOT simulation}

We simulate the red MOT in one dimension with counter-propagating circularly polarized light with components addressing both the $N"$=0, G"=0 and G"=1 manifolds and exciting to available \Dstate{} states. Initial conditions are given by a 1~mK distribution using the parameters of an orange MOT with a trap oscillation frequency of $\omega_{MOT}$ = 2$\pi\times$155 Hz \cite{Hummon2013}. Because the Doppler broadened linewidth at 1~mK is significantly greater than the narrow transition, we model the laser light with multiple frequency components spanning the broadened linewidth, each with saturation parameter $s_0$ \cite{Loftus2004,Loftus2004A}. We use a carrier plus two pairs of sidebands spaced by 2$\pi\times$1~MHz, with the detuning of the highest energy sideband from resonance labeled $\delta_{1}$. Since microwave mixing and vibrational repumping are both fast compared to the optical cycling on this transition, we make the following assumptions about the population distribution in the ground state. Because of the 760~MHz hyperfine splitting, $G"$=0 manifolds are only mixed with other $G"$=0 states, and $G"$=1 manifolds with other $G"$=1 states. So, population in a given $G"$ manifold is weighted by the branching ratio from the \Dstate{} state, which is 1/4 and 3/4 into the $G"$=0 and 1 levels respectively. For a given $G"$ level, the population is evenly distributed among all available $v"$=0, $N"$=0,1,2 states due to the microwave mixing. 
For a two level system, the force from a laser beam frequency component resulting in absorption and followed by spontaneous emission can be written as \cite{Metcalf1999}
\begin{equation*}
F = \frac{\hbar k s_{0}\gamma/2}{1+s_{0}+(2\delta/\gamma)^{2}}
\end{equation*} 
where $k$ is the wavenumber, $s_{0}$ is the saturation parameter, $\delta$ the detuning experienced by the molecule, and $\gamma$ = 1/$\tau$. In a MOT, $\delta$ can be written as $\delta = \delta_{0} -k v +\mu^{'}Ar/\hbar$, where $\delta_{0}$ is the nominal detuning of the laser frequency component from resonance, $v$ the velocity, $A$ the magnetic field gradient, $r$ the molecule's position, and $\mu^{'}$ the differential magnetic moment for the transition. 
To compute the average force felt by an $N"=0$ molecule, the force is averaged over the transitions available to all $N"=0$ sublevels, and summed over both laser beams and all frequency components. 
Since we only optically address $N"$=0 which contains 1/9 of the microwave mixed states, the average force felt by a molecule mixed between $N"$=0,1,2 is $\bar{F}_{total}$=$\bar{F}_{N"=0}$/9. In addition, we simulate heating due to spontaneous emission by adding a random momentum kick when a photon is absorbed. For small detunings, the force reduces to  $\bar{F}_{total}=-\beta v-\kappa r$, where $\beta$ is the damping cofficient and $\kappa$ the spring constant, which together characterize the trap. 

\begin{figure}[h]
\includegraphics[width=\textwidth]{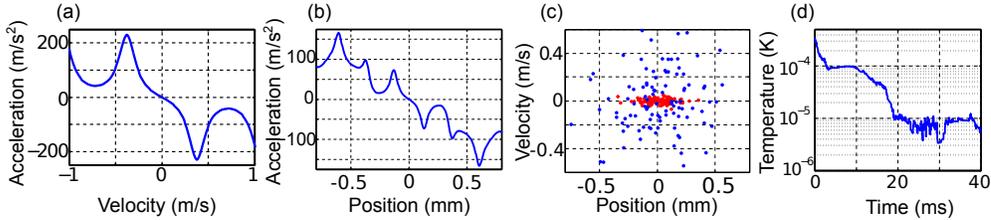}
\caption{Results of simulation. (a) Velocity dependent force due to optical molasses. (b) Position dependent force due to magnetic field gradient for stationary molecules. (c) 1~mK distribution produced from orange MOT (blue) and distribution after red MOT has been on for 20~ms (red). (d) Temperature vs time after red MOT is turned on and the orange MOT is turned off.}
\label{fig:simulation}
\end{figure}

We first model an optical molasses as shown in Figure \ref{fig:simulation}(a) for the parameters $\delta_{1}$ = -1.5$\gamma$ and $s_{0}$ = 3, chosen to provide force to all moving molecules in an initial 1 mK distribution. From the slope around zero velocity, we can extract $\beta$ and so the damping rate $\Gamma_{D}=2\beta/m$ = 950~Hz. Likewise, we extract $\kappa$ from the rate of change of force with position for zero velocity molecules, as shown in Figure \ref{fig:simulation}(b). Under the same parameters, and magnetic field gradient $A$ = 3~G/cm, we find a MOT oscillation frequency $\omega_{MOT} = \sqrt{\kappa/m}$ of 2$\pi\times$33~Hz. The characteristic trapping time is then $\Gamma_D/\omega_{MOT}^2$ = 20 ms \cite{Metcalf1999}. Figures \ref{fig:simulation}(c) and \ref{fig:simulation}(d) show the results of a Monte-Carlo simulation of the narrow-line MOT. By comparing the initial temperature distribution (blue) and after 20~ms (red), we see clear phase space compression. The temperature after 20~ms is less than 10~$\mu$K. The simulation indicates a photon scatter rate $\sim$$10^{4}$ per second, and cooling to \textless10~$\mu$K requires only $\sim$200 photons. A naive worst case extrapolation to three dimensions is yielded by dividing $\kappa$ and $\beta$ by three, giving $\omega_{3D} = 2\pi\times$19~Hz, $\Gamma_{3D} =$317~Hz, and requires $\sim$600~photons to cool. We estimate the $v"=3$ limited MOT lifetime to be \textgreater1~second. We note that for a molecule at rest, the maximum acceleration due to the three dimensional MOT is $\sim$5 times that of gravity, allowing successful trapping. In sum, the simulation indicates that the red MOT is capable of trapping and further cooling the molecules captured in an orange MOT, requiring only the addition of one 690~nm laser with frequency components split by 760~MHz, obtainable with an acousto-optic modulator, to the first stage system, one for the $G"$=0 and the other for the $G"$=1 to \Dstate{} transition.

\section{Conclusion}
We have identified a transition in YO suitable for secondarily cooling a MOT on the \Sstate{} to \Pstate{1} transition by using a narrow-line transition from the \Sstate{} to \Dstate{} state. Two estimates indicate a lifetime of $\tau$ $\approx$1~$\mu$s for the $A'^{2}\Delta$ state, allowing a final temperature of $\sim$10~$\mu$K. We have also determined that the vibrational branching from the \Dstate{} state is favorable enough to not necessitate any more vibrational repump lasers than the first stage MOT. Simulations indicate parameters for the red MOT of a damping rate of 317~Hz and a MOT frequency of 2$\pi\times$19~Hz.

\ack
We thank for Ben Stuhl for the original idea of narrow-line cooling in YO. We acknowledge funding support from ARO, AFOSR (MURI), Gordon and Betty Moore Foundation, NIST, and NSF.

\end{document}